\documentclass[twocolumn,preprintnumbers,amsmath,amssymb]{revtex4}
\usepackage{graphicx}

\begin{document}

\title{Predicting the size of droplets produced through Laplace
pressure induced snap-off}

\author{Solomon Barkley$^{a}$}
\author{Samantha J. Scarfe$^{a}$}
\author{Eric R. Weeks$^{b}$}
\author{Kari Dalnoki-Veress$^{a,c}$}
\email{dalnoki@mcmaster.ca}
\affiliation{$^a$ Department of Physics and Astronomy, McMaster
University, Hamilton, ON, Canada}
\affiliation{$^b$ Department of Physics, Emory University,
Atlanta, GA 30322, USA}
\affiliation{$^{c}$~PCT Lab, UMR CNRLS 7083 Gulliver, ESPCI ParisTech, PSL Research University, Paris, France.}

% & \noindent\large{Solomon Barkley,\textit{$^{a}$} Samantha J.
% Scarfe,\textit{$^{a}$} Eric R. Weeks,\textit{$^{b}$} and Kari
% Dalnoki-Veress $^{\ast}$\textit{$^{ac}$}} \\

\begin{abstract}
Laplace pressure driven snap-off is a technique that is used to produce droplets for emulsions and microfluidics purposes. Previous predictions of droplet size have assumed a quasi-equilibrium low flow limit. We present a simple model to predict droplet sizes over a wide range of flow rates, demonstrating a rich landscape of droplet stability depending on droplet size and growth rate. The model accounts for the easily adjusted experimental parameters of geometry, interfacial tension, and the viscosities of both phases. 
\end{abstract}

\maketitle

%%%MAIN TEXT%%%%
\section{Introduction}
\begin{figure}
\centering
\includegraphics[width=7cm]{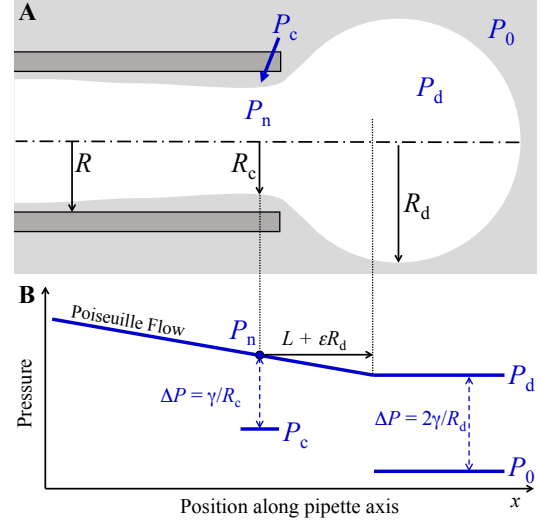}
\caption{A) Schematic prior to droplet snap-off. The inner radius of the cylindrical nozzle $R$, minimum radius of the flowing dispersed phase column within the nozzle $R_\mathrm{c}$, and radius of the growing droplet $R_\mathrm{d}$ define the important length scales. The relevant pressures are the pressure of the bulk continuous phase $P_\mathrm{0}$, pressure of the dispersed phase in the growing droplet $P_\mathrm{d}$, pressure of the dispersed phase in the nozzle $P_\mathrm{n}$, and the pressure of the continuous phase in the collar around the dispersed phase $P_\mathrm{c}$. B) Schematic of these pressures along the length of the pipette.  Snapoff occurs if $P_c < P_0$, which happens as $R_d$ grows, decreasing $P_d$ and the other pressures relative to the fixed pressure $P_0$.}
\label{fig:schematic}
\end{figure}

With the development of numerous techniques to produce micron-scale droplets, there has been a growing interest in exploring the physical mechanisms behind these methods \cite{utada2007dripping,dangla2013physical,garstecki2006formation,sugiura2002characterization, van2010mechanism}. A greater understanding of which parameters are most important to droplet production can permit fine-tuning of existing techniques or modification of systems where droplet production is undesirable \cite{pena2009snap,sugiura2004effect,dangla2013droplet}. Laplace pressure driven snap-off relies upon an instability that forms when a confined dispersed phase is allowed to penetrate into a larger space filled with an immiscible continuous phase (see Fig.~\ref{fig:schematic}A). 
The instability requires no viscous interaction, but is due entirely to changes in the curvature of the interface between the two phases, affecting the Laplace pressure. We have previously found that the resulting droplets are highly monodisperse~\cite{Barkley_TT}, as the system becomes unstable immediately after the protrusion of the dispersed phase reaches a critical size. The snap-off phenomenon has been studied and used to produce droplets in many different geometries \cite{van2010mechanism, dangla2013droplet, Lenormand1983mechanisms, umbanhowar00, sugiura2002characterization,Barkley_TT}. However, the case of ejecting the dispersed phase from a cylindrically symmetric tube is the simplest from a theoretical standpoint \cite{roof1970snap,pena2009snap,Barkley_TT}. 

By neglecting pressure gradients within the dispersed phase, several groups have been able to provide a theoretical framework for droplet sizes in a quasi-static regime \cite{Barkley_TT,dangla2013droplet,van2010mechanism,roof1970snap}. The approximation of a static dispersed phase has been shown to be valid in regimes that are useful for droplet production \cite{Barkley_TT,dangla2013physical,van2010mechanism}. Additionally, it was shown that droplets grow larger before snapping off at higher flow rates, and eventually grow indefinitely for sufficiently large flow rates \cite{Barkley_TT,pena2009snap,dangla2013droplet,sugiura2002effect}. The ability to deliberately vary droplet size without changing the apparatus, for instance by changing the flow rate, represents a useful experimental control variable. Nevertheless, a model which can predict the size of snap-off droplets at flow rates where the quasi-static approximation is no longer valid has remained elusive. Previous studies have examined the effects of changes in geometry, interfacial tension, and viscosity ratio between the two fluids on the droplet production\cite{van2010mechanism, pena2009snap, sugiura2002effect, sugiura2004effect, dangla2013physical, dangla2013droplet}. However, many of these investigations have been largely qualitative in nature.

Here, we present experiments that are compared to a model predicting the size of droplets produced through the snap-off mechanism over a wide range of flow rates. The model provides a ``stability diagram'': regions on a plot of the droplet radius as a function of the flow rate where droplets are stable and grow, contrasted with regions where the droplets are unstable and snap off. The stability diagram elucidates a pathway for manipulation of the system to produce droplets at both sizes and flow rates where snap-off would not normally occur. Furthermore, the model we present accounts for the effects of fluid properties and geometry on the droplet production. The work is carried out in the  cylindrically symmetric geometry, although the ideas are easily extended to the flattened geometries that are common for snap-off droplet production.

\section{Theory}
Snap-off occurs when a dispersed phase is ejected from a nozzle into a reservoir of a continuous phase as shown in Fig.~\ref{fig:schematic}A. The dispersed phase forms a growing spherical droplet that becomes unstable at some critical size and subsequently snaps-off~\cite{Barkley_TT}.  Figure~\ref{fig:schematic}B illustrates schematically the location and relative values of  the relevant pressures, which we now describe. Assuming a quasi-equilibrium, we can define the pressure of the bulk continuous phase $P_0$.   The interior pressure of the droplet of dispersed phase is higher than $P_\mathrm{0}$ due to the droplet's Laplace pressure:
\begin{equation}
P_\mathrm{d}=P_\mathrm{0}+2\gamma/R_\mathrm{d},
\label{eq:laplace}
\end{equation}
where $R_\mathrm{d}$ is the radius of the droplet and $\gamma$ is the interfacial tension between the dispersed and continuous phase (Fig.~\ref{fig:schematic}B). In more complex, non-cylindrical geometries, growth of the droplet is restricted in one or more directions, and the Laplace pressure difference will depend on the dimensions of the confining chamber \cite{roof1970snap,dangla2013droplet,van2010mechanism}. For snap-off to proceed effectively, there must be a wetting layer of the continuous phase coating the interior of the nozzle, which forms a transient collar with a minimum radius, $R_\mathrm{c}$~\cite{Barkley_TT,roof1970snap}. Again, due to the curvature of the interface at the collar, we can write the Laplace pressure difference between the dispersed phase in the nozzle $P_\mathrm{n}$ and the continuous phase in the surrounding collar $P_\mathrm{c}$,
\begin{equation}
P_\mathrm{n}=P_\mathrm{c}+\frac{\gamma}{R_\mathrm{c}}.
\label{eq:collar}
\end{equation}
Here we have made a simplifying assumption consistent with experimental obsevations, that ${R_\mathrm{c}}$ is a much smaller radius than that of the orthogonal curvature.  As the droplet begins its growth, the wetting layer of the continuous phase is thin, and so $R_\mathrm{c} \approx R$.

Figure~\ref{fig:schematic}B also indicates another important feature of the flowing fluid, that there is a pressure gradient, ${dP}/{dx}$, along the cylindrical nozzle.  This pressure gradient is described by Poiseuille flow~\cite{landau1987},
\begin{equation}
\frac{dP}{dx}=\frac{8\eta Q}{\pi R^4},
\label{eq:poiseuille}
\end{equation}
where $\eta$ is the viscosity of the dispersed phase and $Q$ is the volumetric flow rate.  We are concerned with the pressure difference between the dispersed phase at the collar, $P_\mathrm{n}$, and that in the droplet, $P_\mathrm{d}$. In practice, nozzles with circular apertures are undesirable, as spherical droplets block the opening and prevent the reverse flow of the continuous phase that is necessary for snap-off to occur \cite{roof1970snap,Barkley_TT}. Cylindrical nozzles must therefore have an irregular tip shape that is not easily controlled between nozzles. The decrease in pressure from $P_\mathrm{n}$ to $P_\mathrm{d}$ occurs over a length scale with two distinct components. First, each tip contributes a geometric parameter $L$ that is specific to the pipette tip's unique shape, reflecting the distance into the pipette at which the neck of the dispersed phase forms. Second, the pressure must drop to $P_\mathrm{d}$ by some distance into the droplet, which contributes $\epsilon R_\mathrm{d}$, since this dimension scales with the droplet radius. Thus the length scale over which the pressure drops from $P_\mathrm{n}$ to $P_\mathrm{d}$ is $L + \epsilon R_\mathrm{d}$ and with  Eq.~\ref{eq:poiseuille}, we obtain 
\begin{equation}
P_\mathrm{n}=P_\mathrm{d}+\frac{8\eta Q}{\pi R^4}\left(L+\epsilon
R_\mathrm{d}\right).
\label{eq:flowpressure}
\end{equation}

Equations \ref{eq:laplace}, \ref{eq:collar}, and \ref{eq:flowpressure} establish the relationships between each of the pressures depicted depicted in Fig.~\ref{fig:schematic}.  As the droplet grows, the Laplace pressure difference between the droplet interior and exterior decreases via Eq.~\ref{eq:laplace}, and so $P_\mathrm{d}, P_\mathrm{n},$ and $P_\mathrm{c}$ all decrease, relative to the constant $P_\mathrm{0}$.  The snap-off instability develops when $P_\mathrm{c}<P_\mathrm{0}$, causing a reverse flow of the continuous phase from the bulk $P_\mathrm{0}$ into the collar $P_\mathrm{c}$. As the continuous phase flows, the collar constricts the dispersed phase (${R_\mathrm{c}}$ decreases), which further reduces $P_\mathrm{c}$ relative to $P_\mathrm{n}$ according to Eq.~\ref{eq:collar}.  The contracting collar  exacerbates the pressure difference driving reverse flow of the continuous phase, and the collar structure rapidly collapses, releasing the droplet \cite{Barkley_TT,dangla2013physical}.  Using our relations between the pressures and with $R_\mathrm{c} \approx R$, we can rewrite the snap-off condition $P_\mathrm{c}<P_\mathrm{0}$ as:
\begin{equation}
P_\mathrm{0}+\frac{2\gamma}{R_\mathrm{d}}+\frac{8\eta Q}{\pi
R^4}\left(L+\epsilon R_\mathrm{d}\right)-\frac{\gamma}{R} < P_\mathrm{0}.
\label{eq:snapoff}
\end{equation}
When this inequality is satisfied, the collar is unstable and shrinks to release  the droplet.

Previous efforts to predict snap-off droplet size have been restricted to the low flow limit of the dispersed phase \cite{Barkley_TT,dangla2013physical,van2010mechanism,roof1970snap}.  In this case, $Q \approx 0$, $P_\mathrm{d} \approx P_\mathrm{n}$, and Eq.~\ref{eq:snapoff} simplifies to the condition 
\begin{equation}
R_\mathrm{d} > 2R,
\label{eq:quasistatic}
\end{equation}
as we have shown in a previous study\cite{Barkley_TT}.  That is, when $R_\mathrm{d}$ grows to the size of $2R$, the instability begins and the droplet snaps off, and so all droplets formed are of size $2R$ in the low flow limit.

For higher flow rates, we can introduce the non-dimensional variables $\widetilde{R}_\mathrm{d}=\frac{R_\mathrm{d}}{R}$, $\widetilde{L}=\frac{L}{R}$, and $\widetilde{Q}=\frac{8 \eta}{\pi \gamma R^2}Q$.  With these varibles, Eq.~\ref{eq:snapoff} can be rewritten to define unstable droplets as
\begin{equation}
\epsilon \widetilde{Q}\widetilde{R}_\mathrm{d}^2 +\left(\widetilde{L}\widetilde{Q}-1 \right)\widetilde{R}_\mathrm{d}+2 < 0.
\label{eq:nondimensional}
\end{equation}
The expression can be solved to find the values of $\widetilde{R}_\mathrm{d}$ for which the left side equals zero:
\begin{equation}
\widetilde{R}_\mathrm{d}=\frac{1-\widetilde{L}\widetilde{Q}\pm\sqrt{\left(1-\widetilde{L}\widetilde{Q}\right)^2-8\epsilon\widetilde{Q}}}{2\epsilon\widetilde{Q}}.
\label{eq:rofQ}
\end{equation}

\begin{figure}
\centering
\includegraphics[width=9cm]{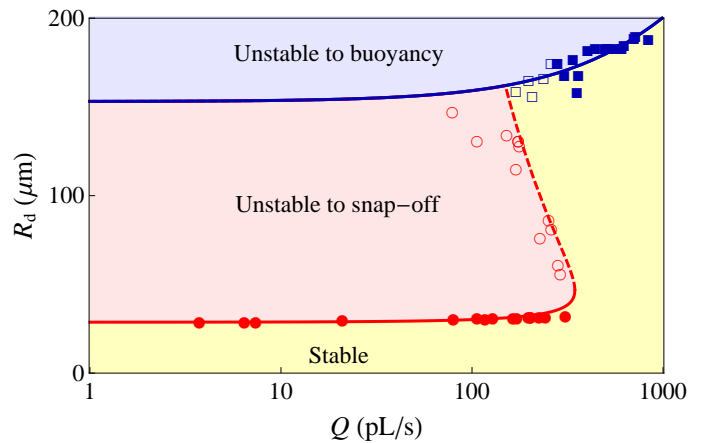}
\caption{Stability diagram for droplets as they grow at a particular flow rate from a single pipette ($R=14.3~\mu$m). Droplets growing at a low constant flow rate undergo snap-off (solid circles) when their radius surpasses the critical size described by Eq.~\ref{eq:rofQ} (lower solid line, with $\eta=123$~mPa$\cdot$s, $\gamma=12$~mN/m). Droplets growing at a high constant flow rate break away from the pipette due to buoyancy (solid squares), when they reach the size specified by Eq.~\ref{eq:buoyancy} (upper solid line, with $V_\mathrm{b}=15~$nL, $\tau=19~$s). Droplets produced with a flow rate that is not constant and decreasing can be made to snap-off at intermediate sizes (open symbols) after crossing the dashed line (Eq.~\ref{eq:rofQ}).}
\label{fig:phasediagram}
\end{figure}

Eq.~\ref{eq:rofQ} defines a critical non-dimensional droplet radius $\widetilde{R}_\mathrm{d}$ for a given rescaled flow rate $\widetilde{Q}$. Beyond the critical radius, the continuous phase invades the nozzle containing the dispersed phase, causing the droplet to snap-off. Eq.~\ref{eq:rofQ} can thus be interpreted as follows: For a given flow rate, the droplet grows, while still attached by the collar to the dispersed phase. Eventually the droplet grows to the critical radius given by Eq.~\ref{eq:rofQ} at which point the collar pinches off and the droplet is released into the continuous medium. Thus Eq.~\ref{eq:rofQ} defines the boundary between stable and unstable droplets in a parameter space of $\widetilde{R}_\mathrm{d}$ and $\widetilde{Q}$ as shown by the lower solid line in Fig.~\ref{fig:phasediagram}.

The above explanation assumes that the negative square root is taken in Eq.~\ref{eq:rofQ}. If the positive square root is taken instead, Eq.~\ref{eq:rofQ} defines a minimum stable droplet size as a function of flow rate, shown as the dashed line in Fig.~\ref{fig:phasediagram}. Droplets larger than this critical size are stable to continue growing indefinitely as the $\epsilon \widetilde{Q}\widetilde{R}_\mathrm{d}^2$ term dominates Eq.~\ref{eq:nondimensional}.  In this situation, the distance $\epsilon R_\mathrm{d}$ is large.  Thus, the pressure gradient necessary to drive the flow into the large droplet ensures that $P_\mathrm{n}$, and thus $P_\mathrm{c}$, stay large and $P_\mathrm{c} > P_0$ even for arbitrarily large droplets.

In reality, no droplets are able to grow indefinitely, as large droplets break away from the pipette due to their buoyancy. It is important to note that this separation represents a different mechanism than the Laplace-pressure driven snap-off.  Here, the buoyant force overcomes the interfacial tension, similar to droplets forming on a leaky ceiling \cite{weisskopf86}.  The size at which buoyant forces are sufficient to separate the droplet from the pipette depends on both the shape and orientation of the tip, as well as the flow rate. We assume that for a particular pipette, there is some critical droplet volume $V_\mathrm{b}$ beyond which the droplet is unstable to buoyancy.  However, droplets take some fixed amount of time $\tau$ to detach from the pipette and continue to grow at the same flow rate during this time. The volume of droplets that break away from the pipette is therefore expressed simply as $V_\mathrm{d}=V_\mathrm{b}+Q\tau$. The droplet radius at buoyant separation is then given by:
\begin{equation}
R_\mathrm{d}=\left(\frac{3}{4\pi}\left(V_\mathrm{b}+Q\tau\right)\right)^{\frac{1}{3}},
\label{eq:buoyancy}
\end{equation}
which is plotted as the upper solid curve in
Fig.~\ref{fig:phasediagram}.  While $V_b$ depends on several
details of the pipette tip geometry, we note that it should scale
as
\begin{equation}
V_\mathrm{b} \sim \left( \frac{\gamma}{\Delta \rho g} \right)^{3/2},
\label{eq:vb}
\end{equation}
suggesting that larger droplets are possible by increasing the
surface tension or better matching the densities of the two
fluids.

\section{Experiment}
Cylindrical nozzles were created by stretching capillary tubes (World Precision Instruments, USA) with a pipette puller (Narishige, Japan). Initial capillary inner diameter was $540~\mu$m and final pipette inner diameter was on the order of $\approx 10~\mu$m. Pipette ends were clipped with tweezers after stretching to produce a jagged tip shape unique to each pipette. As discussed above (see also~\cite{Barkley_TT}), jagged pipettes were found to perform optimally in comparison to perfectly flat ends that allow the droplet to seal the tip, thus preventing the reverse flow that results in snap-off. In fact, the pipettes are not perfectly cylindrical, but taper gradually from diameter $\approx 100~\mu$m to $\approx 10~\mu$m over several centimetres, and so pipettes with larger openings can be produced by breaking the pipettes further from the tip. Unless otherwise stated, the dispersed phase was mineral oil and the continuous phase was water with 1\%  (by weight) sodium dodecyl sulphate (SDS) added to stabilize droplets against fusion. Droplets were produced by ejecting oil out of a pipette into a chamber filled with the water phase, taking care that the pipette tip was far from any chamber boundaries. The flow rate was controlled by maintaining a connected oil reservoir at a constant pressure, either by adjusting its height or by clamping a syringe. This procedure results in a constant flow rate at the pipette tip as long as the volume of oil leaving the pipette is negligible compared to the oil reservoir. Droplets were imaged from below at up to 50 frames per second. 

The viscosity of the oil phase was decreased for some experiments by mixing dodecane with the mineral oil in concentrations up to $50\%$ by weight. The viscosity of the water phase was increased for some experiments by adding glycerol at concentrations up to $60\%$, with the  SDS concentration maintained at $1\%$. The interfacial tension between the two phases was increased for some experiments by reducing the concentration of SDS in the water phase, to a minimum of $0.05\%$. The viscosity of mineral oil was measured by allowing a $225~\mu$m diameter polystyrene bead (Duke Scientific, USA) to slowly fall through a column of oil. A measurement of its terminal velocity was used to calculate viscosity through Stokes' law. The viscosity measurements were repeated with the mineral oil and dodecane mixtures, and its results compared to predictions of a method to calculate viscosity of hydrocarbon blends~\cite{ASTM2006}.  

For certain experiments, it was necessary to reduce the oil flow rate during droplet growth. By doing so, the trajectory of a droplet's growth through the parameter space depicted in fig.~\ref{fig:phasediagram} is altered from a simple vertical line. To accomplish this, the oil reservoir was slowly and continuously lowered once the droplet had grown to an intermediate size. The oil reservoir was fixed at its new height upon snap-off of the growing droplet. For droplets growing at a constant flow rate, total droplet volume released in a timed experiment was used to determine the flow rate. In order to measure the flow rate at snap-off for droplets growing at a decreasing flow rate, a number of additional droplets were allowed to form after fixing the reservoir height, and the flow rate of these droplets was measured.

\section{Results \& Discussion}

For a given pipette, droplets can be produced at many different flow rates. At low flow rates, droplet size is independent of flow rate over orders of magnitude changes to $Q$, as predicted by Eq.~\ref{eq:quasistatic} (solid circles in Fig.~\ref{fig:phasediagram}). At higher flow rates, droplet size at snap-off increases with flow rate, until the flow rate reaches some critical value. This observation of a critical value agrees well with previous results \cite{Barkley_TT,sugiura2002effect}. For sufficiently large $\widetilde{Q}$, there is no droplet size $\widetilde{R}_\mathrm{d}$ that can satisfy Eq.~\ref{eq:nondimensional} and droplets would be expected to grow indefinitely \cite{pena2009snap,Barkley_TT}. In practice, droplets produced beyond the critical flow rate eventually break away from the pipette due to their buoyancy (solid squares in Fig.~\ref{fig:phasediagram}). Droplet radius at the time of separation due to buoyancy is described by Eq.~\ref{eq:buoyancy} and is plotted as the upper curve in Fig.~\ref{fig:phasediagram}, which is a fit to the square data points ($V_\mathrm{b}=15~$nL, $\tau=19~$s).

Droplets growing at a constant flow rate can be interpreted as an upwards vertical trajectory in the stable region of Fig.~\ref{fig:phasediagram}. Droplets will break away from the pipette either upon snap-off at low flow rates (solid circles in Fig.~\ref{fig:phasediagram}), or due to buoyancy at high flow rates (solid squares). For many pipettes, the droplets produced through these two mechanisms have very different sizes and it is not possible to produce droplets of intermediate size at any constant flow rate. The snap-off droplets produced for different constant flow rates agree with the prediction of Eq.~\ref{eq:rofQ} if the negative square root is taken (lower solid curve in Fig.~\ref{fig:phasediagram}, with $\widetilde{L}=-5.2$ and $\epsilon=4.3$). 

Since droplets grow (i.e.  $R_\mathrm{d}$ increases for positive $Q$), it is only possible to travel upwards in the $R_\mathrm{d}$--$Q$ stability diagram shown in Fig.~\ref{fig:phasediagram}. The inverted large-droplet branch of the snap-off curve (dashed line in Fig.~\ref{fig:phasediagram}) is not normally accessible, as it would require passing from an unstable droplet (which cannot remain attached to the pipette) to a stable one. However, individual droplets can be produced at continuously decreasing flow rates, represented as a diagonal trajectory towards the upper left in Fig.~\ref{fig:phasediagram}. When these droplets fulfil an alternate (root) solution to the snap-off condition of Eq.~\ref{eq:rofQ}, they cross the dashed portion of the snap-off stability curve in Fig.~\ref{fig:phasediagram} from right to left. This causes the droplets to become unstable and they subsequently snap-off, represented by the open circles in Fig.~\ref{fig:phasediagram}. As it is not possible to measure the instantaneous flow rate of these droplets, the flow rate of smaller droplets produced immediately afterwards at the same, fixed flow rate is measured instead. Because these subsequent droplets may have actually been produced at a slightly lower flow rate, the flow rate thus obtained provides an underestimate for the flow rate at snap-off of the earlier droplet.

Snap-off of droplets at decreasing flow rates provides a mechanism to produce intermediate sized droplets that are not normally accessible with a particular pipette. However, only individual droplets can be produced in this manner. While there is little practical value of this approach to droplet production, such experiments provided an excellent verification of the model. The presence of this typically inaccessible branch of the snap-off curve confirms that at intermediate flow rates, both small and large droplets are permitted to grow but medium droplets are unstable to snap-off. We reiterate that the stable portion of the $R_\mathrm{d}$--$Q$ diagram corresponding to large droplets is not accessible at a constant flow rate, since this would require droplets to pass through unstable intermediate sizes. 

\begin{figure}
\centering
\includegraphics[width=9cm]{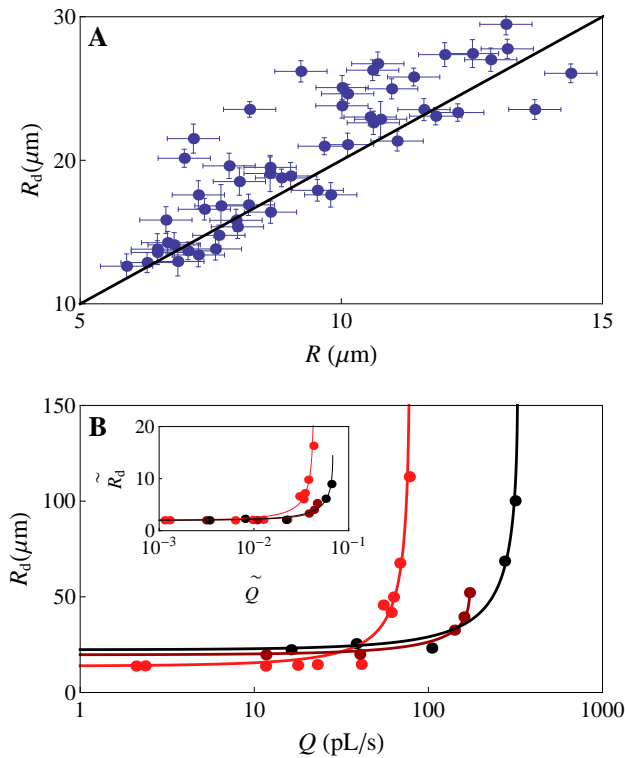}
\caption{Effect of pipette dimensions on droplet radius ($\eta=123$~mPa$\cdot$s, $\gamma=12$~mN/m). A) Minimum observed radius of droplets for many different pipette radii, showing good agreement with prediction for the low-flow limit (black line). B) Snap-off droplet size for increasing flow rate through three different pipettes of radii $6.9~\mu$m, $9.9~\mu$m,$11.2~\mu$m. Curves are fits to Eq.~\ref{eq:rofQ}, returning values of $\widetilde{L}=\{20.2, 5.3, 10.7\}$ and $\epsilon = \{0.06, 1.5, 0.14\}$. Inset presents the same data in terms of scaled droplet radius and flow rate.}
\label{fig:changeR}
\end{figure}

One of the most straightforward methods to produce snap-off droplets of a different size is to alter the chamber dimensions~\cite{Barkley_TT, sugiura2002effect, dangla2013physical, dangla2013droplet}. In the case of a cylindrical nozzle ejecting a dispersed phase into an unconfined bulk continuous phase, the only dimension in the system is the radius of the nozzle, $R$. In the low flow limit, Eq.~\ref{eq:quasistatic} predicts droplet size $R_\mathrm{d}=2R$ or $\widetilde{R}_\mathrm{d}=2$ (see also reference~\cite{Barkley_TT}). This prediction is confirmed by the data presented in Fig.~\ref{fig:changeR}A, which plots the minimum droplet size observed at low flow rates as a function of the inner pipette radius for many pipettes, along with the prediction from Eq.~\ref{eq:quasistatic} (black line). Having established the validity of $R_\mathrm{d}=2R$, for the remainder of this study we obtain $R$ from $R_\mathrm{d}$ because of the accuracy and relative simplicity of that measurement in contrast with the direct measurement of the inner radius of the pipette.  At higher flow rates, the dependence of $\widetilde{R}_\mathrm{d}$ on $R$ is non-trivial, as both $\widetilde{L}$ and $\widetilde{Q}$ in Eq.~\ref{eq:rofQ} are normalized by $R$ and $R^2$, respectively. Fig.~\ref{fig:changeR}B shows data from three pipettes with different radii. For each pipette, the characteristic curve of droplet radius $R_\mathrm{d}$ is plotted as a function of volumetric flow rate $Q$. The same data is plotted in terms of the non-dimensionalized variables $\widetilde{R}_\mathrm{d}$ and $\widetilde{Q}$ in the inset of Fig.~\ref{fig:changeR}B, demonstrating good collapse in the vertical direction (collapse is not expected in the horizontal direction, even if pipettes shared values of $\widetilde{L}$ and $\epsilon$, since $\widetilde{Q}$ appears with each of these parameters separately in Eq.\ref{eq:rofQ} and they exhibit dissimilar dependence on $R$). All datasets were fit to Eq.~\ref{eq:rofQ}, with oil viscosity measured as $\eta=123 \pm 4$~mPa$\cdot$s and interfacial tension taken as $\gamma=12$~mN/m based on related literature~\cite{campanelli1997effect,rehfeld1967adsorption,sugiura2004effect}. Both $\widetilde{L}$ and $\epsilon$ are expected to depend on the precise tip shape, and vary between pipettes (for the three pipettes used in Fig.~\ref{fig:changeR}, $\widetilde{L}=\{20.2, 5.3, 10.7\}$ and $\epsilon = \{0.06, 1.5, 0.14\}$). 

\begin{figure}
\centering
\includegraphics[width=9cm]{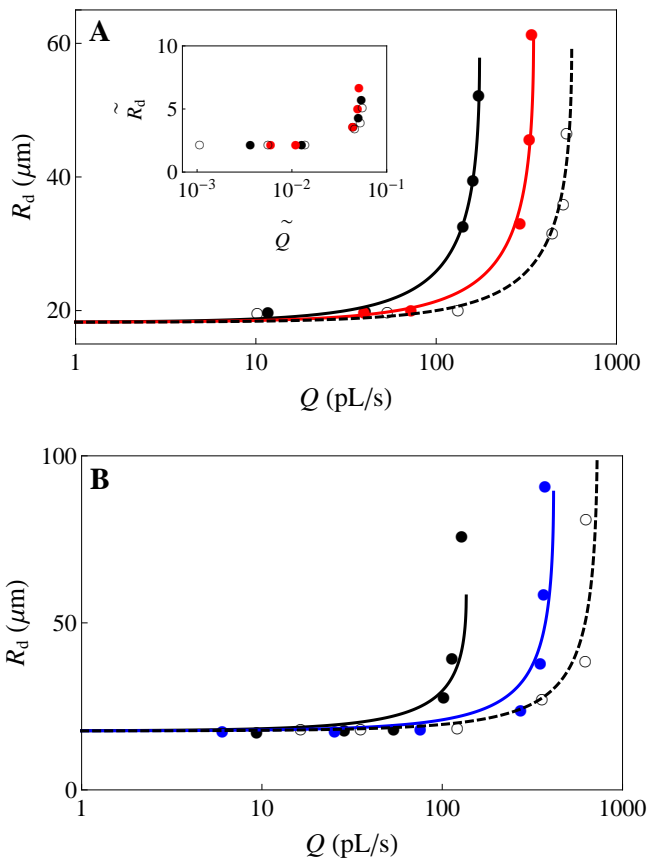}
\caption{A) Droplets sizes produced from a single pipette ($R=9.1~\mu$m) at increasing flow rates along with fits to Eq.~\ref{eq:rofQ} for SDS concentrations of 1\% (black circles and solid curve), 0.01\% (lighter filled circles and solid curve), and 0.005 \% (open circles and dashed curve). Interfacial tensions were 12 mN/m, 25 mN/m, and 36 mN/m, respectively. Viscosity was held fixed at $\eta=123$~mPa$\cdot$s. Inset presents the same data plotted in terms of non-dimensional units. B) Droplet sizes produced from a single pipette at increasing flow rates along with fits to Eq.~\ref{eq:rofQ} for oil viscosities of 123 mPa$\cdot$s (black circles and solid curve), 20 mPa$\cdot$s (lighter filled circles and solid curve) and 6.2 mPa$\cdot$s (open circles and dashed curve). Interfacial tension was constant at $\gamma=12$~mN/m.}
\label{fig:changeoil}
\end{figure}

The snap-off process is driven by the interfacial tension between
the two phases. However, the actual value of $\gamma$ does not
affect droplet production in the low flow limit, since it is present
in each term of Eq.~\ref{eq:quasistatic} (this was also observed
in reference~\cite{sugiura2004effect}). In contrast, at higher flow
rates, interfacial tension matters as it influences the non-dimensionalized flow
rate $\widetilde{Q}\propto Q/\gamma$. The bare interfacial tension
between mineral oil and water is large ($\gamma=48$~mN/m),
and is lowered in this system through the addition of SDS
surfactant to the water phase~\cite{campanelli1997effect}. By
reducing the concentration of SDS below the critical
micelle concentration, it is possible to increase
the interfacial tension relative to the solution with $1\% $
SDS~\cite{sugiura2004effect,campanelli1997effect,rehfeld1967adsorption}.
Fig.~\ref{fig:changeoil}A shows the snap-off droplet radius for
three different interfacial tensions at increasing flow rates for
the same pipette. Since snap-off is driven by interfacial tension,
droplets snap-off sooner at increased interfacial tension for a
given flow rate. Smaller droplets are produced as a result, which
can be clearly seen in Fig.~\ref{fig:changeoil}A, where the curves
of higher interfacial tension fall below those of lower interfacial
tension. Based on previous experiments with different hydrocarbons,
we estimate the interfacial tension of mineral oil and water at SDS
concentrations of 1\%, 0.1\%, and 0.05\% to be 12 mN/m, 25 mN/m,
and 36 mN/m, respectively~\cite{rehfeld1967adsorption}. Interfacial
tension is accounted for in the non-dimensionalization of $Q$, and is not expected to influence $\widetilde{L}$ or $\epsilon$, which depend only on the pipette tip shape. Upon fitting to Eq.~\ref{eq:rofQ}, $\widetilde{L}=7.2\pm0.5$, and $\epsilon=0.84\pm0.06$ for all SDS concentrations(curves in Fig.\ref{fig:changeoil}A). To further illustrate the role of interfacial tension, the inset of Fig.~\ref{fig:changeoil}A presents the same data in terms of non-dimensional variables $\widetilde{R}_\mathrm{d}$
and $\widetilde{Q}$, calculated using the estimated values of
$\gamma$. The collapsed datasets confirm the role of interfacial
tension in rescaling the flow rate.

Previous snap-off studies have investigated the role of the viscosity ratio of dispersed phase to continuous phase, $\eta/\eta_\mathrm{w}$~\cite{pena2009snap,van2010mechanism}. As with interfacial tension, neither viscosity value is important in the low-flow limit (Eq.~\ref{eq:quasistatic})~\cite{dangla2013droplet, van2010mechanism}. At higher flow rates, Eq.~\ref{eq:snapoff} predicts a dependence of droplet size on the viscosity of the dispersed phase only, which results from the Poiseuille flow through the nozzle opening (Eq.~\ref{eq:poiseuille}). This dependence is accounted for in the scaling of $\widetilde{Q}\propto\eta Q$, as in the case of interfacial tension, above. The viscosity of the oil phase was reduced by adding dodecane to the mineral oil. Although this change also has some impact on other properties of the oil, such as interfacial tension and density, the change to viscosity is expected to dominate for the purposes of snap-off droplet production. Viscosities of the $25\%$ and $50\%$ solutions of dodecane in mineral oil were calculated as $20 \pm 1$~mPa$\cdot$s and $6.2\pm 0.3$~mPa$\cdot$s, respectively~\cite{caudwell2004viscosity,ASTM2006}. Measurements of these viscosities through Stokes' law agreed with the calculated values. Fig.~\ref{fig:changeoil}B plots snap-off droplet radius $R_\mathrm{d}$ as a function of flow rate $Q$ for three oil phase viscosities ejected from the same pipette. According to Eq.~\ref{eq:poiseuille}, a lower viscosity is associated with a smaller pressure gradient for the same flow rate. A lower viscosity thus increases the relative importance of the Laplace pressure terms in Eq.\ref{eq:snapoff} and causes snap-off to occur at flow rates where it otherwise would not. As with interfacial tension, viscosity of the dispersed phase can be seen as an adjustment to non-dimensionalized flow rate $\widetilde{Q}$. 

\begin{figure}[t]
\centering
\includegraphics[width=9cm]{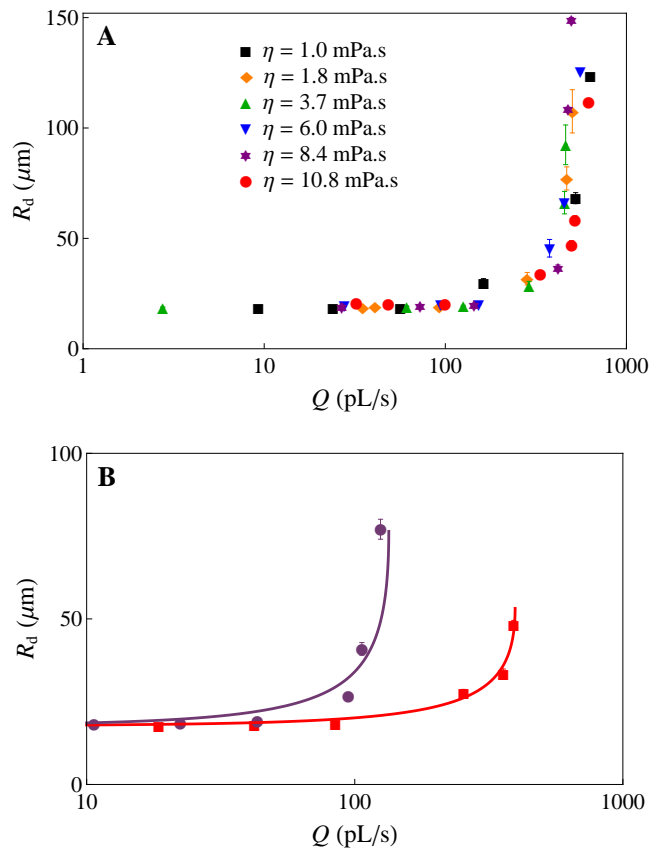}
\caption{A) Snap-off droplet sizes produced from a single pipette ($R=9.1~\mu$m) at increasing flow rates for several viscosities of the continuous phase, with interfacial tension and dispersed phase viscosity held constant at $\eta=123$~mPa$\cdot$s and $\gamma=12$~mN/m. B) Snap-off droplet sizes produced from a single pipette ($R=8.8~\mu$m) for two different liquid pairs with the same viscosity ratio. The viscosity pairs are $\eta=20~$mPa$\cdot$s in $\eta_w=1.0~$mPa$\cdot$s (squares) and $\eta=123~$mPa$\cdot$s in $\eta_w=6.0~$mPa$\cdot$s (circles). Curves are fits to Eq.~\ref{eq:rofQ}}
\label{fig:changewater}
\end{figure}

The theory presented here to describe snap-off (Eq.~\ref{eq:snapoff}) predicts only the onset of droplet instability. It is assumed that the continuous phase invades the nozzle immediately upon fulfilling the snap-off condition and so the threshold for droplet instability is the same as that for actual snap-off. It is for this reason that the theory fails for a perfectly circular opening, since the continuous phase is unable to flow into the nozzle~\cite{Barkley_TT,roof1970snap}. This assumption must also fail in the limit of a highly viscous continuous phase. In order to investigate the role of continuous phase viscosity on snap-off droplet production, glycerol was added to the water phase. As with the addition of dodecane to the mineral oil, addition of glycerol to water is expected to alter interfacial tension and density, but the increased viscosity is expected to dominate changes to the snap-off behaviour.  The viscosity of the water phase does not affect the size of droplets produced from a single pipette at any flow rate, as seen in Fig.~\ref{fig:changewater}A. 
In order to obtain a higher viscosity of the water phase relative to the oil phase, a $50\%$ solution of dodecane in mineral oil was used as the dispersed phase for this experiment, with a calculated and measured viscosity of $\eta=6.2\pm 0.3$~mPa$\cdot$s. The viscosity of the continuous phase varied between $\eta_\mathrm{w}=1.0$~mPa$\cdot$s ($0\%$ glycerol) and $\eta_\mathrm{w}=10.8$~mPa$\cdot$s ($60\%$ glycerol)\cite{segur1951viscosity}.  The experiment was also conducted without addition of dodecane to the dispersed phase with similar results. The independence of snap-off on continuous phase viscosity provides justification for using a stability condition as a snap-off condition directly. Furthermore, the viscosity ratio has previously been assumed to be an important parameter for snap-off production, as is the case for other droplet production techniques~\cite{van2010mechanism,pena2009snap,sugiura2002characterization,utada2007dripping}. However, our results suggest that over the range of viscosities tested here, the viscosity of the dispersed phase dominates the snap-off criterion rather than the ratio of viscosities. To emphasize this point, Fig.~\ref{fig:changewater}B presents two datasets from the same pipette with the same viscosity ratio, $\eta/\eta_\mathrm{w}=20$, but different viscosity pairs. The lack of overlap between these two datasets provides direct evidence that the viscosity ratio is not the correct parameter to consider in the context of snap-off droplet production.

\section{Conclusions}
We have presented a model describing snap-off droplet size over a wide range of flow rates, based on simple arguments of Laplace pressure and Poiseuille flow. Pipette dimensions, interfacial tension, and dispersed phase viscosity are all accounted for. Adjustment of these parameters increases the versatility of droplet production, specifically by increasing the range of flow rates over which droplets can be produced. Interestingly, the continuous phase viscosity has no impact on snap-off over the entire range tested, including cases where the continuous phase is more viscous than the dispersed phase. Although our model was developed for a cylindrical geometry, it should be applicable to all droplet snap-off geometries with appropriate adjustments. The model predicts a rich and unexpected stability dependence on droplet size and flow rate, which was confirmed experimentally by producing droplets with radii that are normally forbidden for a particular pipette.
%%%END OF MAIN TEXT%%%

%\begin{acknowledgement}
Financial Support for this work was provided by National Sciences
and Engineering Research Council (NSERC).  The work of E.R.W. was
supported by the National Science Foundation under Grant No.
CBET-1336401.
%\end{acknowledgement}


%%%REFERENCES%%%

%\bibliography{snap_off_biblio}
%\bibliographystyle{rsc} %the RSC's .bst file

\providecommand*{\mcitethebibliography}{\thebibliography}
\csname @ifundefined\endcsname{endmcitethebibliography}
{\let\endmcitethebibliography\endthebibliography}{}

\end{document}